\documentclass[a4paper,12pt]{article}

\usepackage[super,sort]{cite}
\usepackage{graphicx}
\usepackage{hyperref}
\hypersetup{
     colorlinks = true,
     citecolor  = cyan,
     linkcolor  = magenta
}
\usepackage[usenames,dvipsnames]{xcolor}
\usepackage{amsmath}

\newcommand{\RevisedText}{\textcolor{black}}

\begin{document}

\title{Temperature-Distance Relations in Casimir Physics}

\author{
Mathias Bostr{\"o}m$^{1,2,3}$, 
A. Gholamhosseinian$^4$, 
J. J. Marchetta$^1$, \\
R. W. Corkery$^5$, 
I. Brevik$^6$
}

\date{} 

\maketitle

\begin{center}
\textit{
$^1$Centre of Excellence ENSEMBLE3 Sp. z o. o., Wolczynska Str. 133, \\ 01-919 Warsaw, Poland \\
$^2$Chemical and Biological Systems Simulation Lab, Centre of New Technologies, \\ University of Warsaw, Banacha 2C, 02-097 Warsaw, Poland \\
$^3$Email: mathias.bostrom@ensemble3.eu \\
$^4$Department of Physics, Ferdowsi University of Mashhad, Mashhad, Iran \\
$^5$Surface and Corrosion Science, Department of Chemistry, \\ KTH Royal Institute of Technology, SE 100 44 Stockholm, Sweden \\
$^6$Department of Energy and Process Engineering, \\ Norwegian University of Science and Technology, NO-7491 Trondheim, Norway
}
\end{center}

\begin{abstract}
The Casimir-Lifshitz force arises from thermal and quantum mechanical fluctuations between classical bodies and becomes significant below the micron scale. We explore temperature-distance relations based on the concepts of Wick and Bohr arising from energy-time uncertainty relations. We show that temperature-distance relations similar to those arising from the uncertainty principle are found in various Casimir interactions, with an exact relation occurring in the low-temperature regime when the zero point energy contribution cancels the thermal radiation pressure contribution between two plates.
\end{abstract}

\newpage
\section{Introduction}
\label{Intro}
\par 
 The modern understanding of intermolecular interactions, including van der Waals (dispersion) and Casimir forces, has its origins in historical and ancient observations\,\cite{Casi}. {\RevisedText{Arguably one of the great leaps towards our current understanding was that made by S.C. Wang\,\cite{wang1928problem}, who in 1927 used perturbation theory to solve the Schrödinger equation (of “the new quantum mechanics”) for two hydrogen atoms at large separation, including the interactions of their respective constituent electrons and protons. } By 1961, Lifshitz and co-authors presented a complete general theory of intermolecular dispersion forces for real materials\,\cite{Dzya}, and an alternate derivation was later reported by Parsegian and Ninham using semi-classical electrodynamics theory\cite{ParsegianNinham1969}. 
 Despite great efforts and successes in understanding intermolecular forces, fundamental questions remain unresolved (e.g. the Drude-plasma controversy for finite temperature Casimir forces between metal plates\,\cite{Bost2000,Bord}). 
Our work highlights a curious occurrence where the transition between zero and finite temperature Casimir interactions takes on similar if not, within certain limits, the same distance-temperature relation that can be seen from the energy-time uncertainty relation. This connection supplements prior work heuristically linking the zero-temperature Casimir effect to the energy-time uncertainly relation\cite{gine2018casimir}, and its extension to space times with a minimum length scale\cite{blasone2020heuristic}. 
 We also comment on a temperature-distance relationship arising from the substitution of the Wick relation into the thermal wavelength and similarities between nuclear binding energies and the Casimir effect of perfect conducting plates at femtometer separations with an intermediary electron-positron plasma.

\begin{figure}[!h]
  \centering
  \includegraphics[width=0.7\columnwidth]{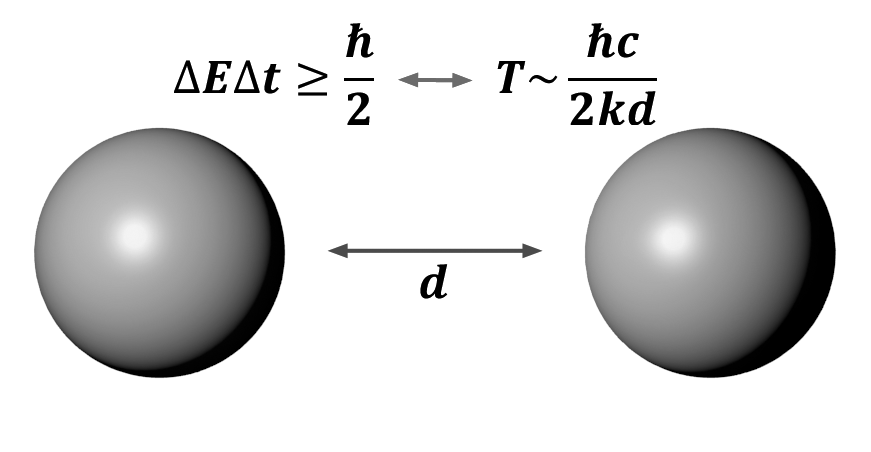}
  \caption{\label{scheme} A schematic figure for two particles that can be nucleons, atoms, molecules, nanoparticles, or macroscopically large metallic surfaces. We derive a link between uncertainties in temperature and distance from the \RevisedText{uncertainty relation}. Similar relations come out from Casimir theory and the theory for a high-temperature Bose-Einstein condensate.}
\end{figure}

\section{Materials and Methods}
\label{MaterialMethods}

\subsection{Concepts}
This section outlines the theoretical foundations and key concepts underlying our study and establishes the framework for investigating temperature-distance relations across different scales of physics.
We begin by reviewing the current understanding of the connections between the Heisenberg uncertainty principle, the quantum harmonic oscillator, and the theory of dispersion forces. Our discussion builds on the pioneering work regarding the theory of dispersion forces in semi-classical electrodynamics, led by Barry W. Ninham and Adrian Parsegian, starting in 1969\cite{ParsegianNinham1969} and supported by highly accurate experimental evidence\,\cite{AndSab,Lamo1997,Mohid}.
Here we explore how Casimir physics can be combined to obtain similar expressions to Wick and Bohrs' temperature-distance uncertainty relations.

\subsection{Heisenberg's Uncertainty Principle and the Theory of Quantum Harmonic Oscillator}
\label{HeisenbergOsc}
 The Heisenberg uncertainty principle arises from the postulates of quantum mechanics and describes
deviations from deterministic classical phase-space trajectories. The uncertainty relation for the canonical position operator  \(\hat{x}\) and momentum operator \(\hat{p}\) \(([\hat{x}, \hat{p}] = i\hbar)\)), which are conjugate pairs in phase space is given by, $(\Delta \hat{x})^2 (\Delta \hat{p})^2 \geq \hbar^2/{4}$,
where $\Delta \hat{x}$ and $\Delta \hat{p}$ represent the standard deviations of position and momentum, respectively, and $\hbar$ is the reduced Planck constant.
The initial uncertainty relation was limited to the operators \(\hat{x}\) and \(\hat{p}\). It was later generalized to any pair of Hermitian operators for pure states\,\cite{schrodinger1930ber}. 
The quantum harmonic oscillator Hamiltonian in terms of position and momentum operators is given by\,\cite{EnergyTemperatureNatureCommun2018},
\begin{equation}
\hat{H} = \frac{\hat{p}^2}{2m} + \frac{1}{2} m \omega^2 \hat{x}^2
\end{equation}
where \(m\) is the mass of the oscillator and \(\omega\) is the angular frequency. This Hamiltonian can be rewritten using the creation (\(a^\dagger\)) and annihilation (\(a\)) operators,
\begin{equation}
{H} = \hbar \omega \left(a^\dagger a + \frac{1}{2}\right).
\label{AydaHamiltonian}
\end{equation}
These operators act on the energy eigenstates \(|n\rangle\), where \(n\) represents the number of excitations, $a^\dagger |n\rangle = \sqrt{n+1} \, |n+1\rangle$ , and $a |n\rangle = \sqrt{n} \, |n-1\rangle$. 
The ground state \(|0\rangle\) is defined by \(a |0\rangle = 0\), and all higher energy states can be generated by repeated application of the creation operator.
As discussed by Wick\,\cite{Wick},  the energy-time uncertainty principle can be applied in deriving a relation between energy and distance $\Delta E \sim {\hbar c}/{2 d}$ which ”gives the distance as the
limit up to which virtual transitions can make themselves felt without contradicting
the energy principle”\cite{Wick}. This describes the range of interaction involving fields with real excitations of mass m and $\Delta E$ is substituted for the rest energy needed for particle creation, $mc^{2}\sim {\hbar c}/{2 d}$.
{\RevisedText {As discussed by Uffink and van Lith\,\cite{Uffink1999} and others \,\cite{Bohr1985,EnergyTemperatureNatureCommun2018}, the uncertainty relations are true milestones in the history of quantum mechanics. In some early writings, attributed\,\cite{Uffink1999,EnergyTemperatureNatureCommun2018} to Bohr\,\cite{Bohr1985}, the possibility of a complementary relationship in classical physics was proposed, particularly between energy and temperature. To assign a definite temperature to a physical system, it must be thermally contacted with a large reservoir acting as a heat bath. The contacted system freely exchanges its energy with the heat bath, the energy fluctuation during the exchange equating to uncertainty in its energy. Upon decoupling from the heat bath, the energy can then be definitely assigned, but then the temperature cannot and so is uncertain. Therefore, as noted in the past\,\cite{Bohr1985,Uffink1999,EnergyTemperatureNatureCommun2018}, just as one finds a symbolic complementary quantum uncertainty relation, one expects to obtain an uncertainty relation for energy and temperature. Dimensional analysis suggests that such a relation is of the form $\Delta U \Delta(1/T) \leq k$. Hence, following in the footsteps of past explorations we propose a relationship linking the classical energy-temperature relation\,\cite{Bohr1985,Uffink1999,EnergyTemperatureNatureCommun2018} with the quantum energy-distance relation proposed by Wick\,\cite{Wick}: $\Delta (kT)\sim \Delta E  \sim {\hbar c}/{2 d}$. 
Our assumptions are supported by alternative derivations given in the past using the same relation between time and temperature $(\Delta t=\hbar/\Delta(kT))$\,\cite{10.1119/1.3194050}. Notably, if these assumptions hold, then distances around 1\,femtometer or less correspond to extremely high-temperature uncertainties of the order $10^{12}$\,K. This temperature range is large enough to potentially generate a quark-gluon plasma at nuclear length scales\,\cite{quarkgluonNature2007}. However, there has been some caution urged against the potential generation of particles from uncertainty relation\,\cite{Roberts_2020}.
From the above discussions,} we propose a relation ($\Delta (kT)\sim \alpha {\hbar c}/{2 d}$),  appear in the Casimir effect, where $\alpha\sim 1$ in all the examples we consider. The prefactor, $\alpha$, equals unity in two specific examples: first when using a \RevisedText{uncertainty relation} and second when the zero-temperature Casimir and repulsive black body radiation terms cancel. Actually, in Sec.\,\ref{TempDistOscillator} a third heuristic derivation leads to the same result.

\subsection{Casimir-Lifshitz theory and the oscillator free energy}

The interaction free energy, under the Casimir-Lifshitz formalism, originates from changes in the electromagnetic field oscillator free energy due to the presence of boundaries\,\cite{NinhamParsegianWeiss1970},

\begin{equation}
   G(d)=\sum_{\omega_j} [g(\omega_j;d)-g(\omega_j;\infty)].
   \label{GibbsOscillator}
\end{equation}
The free energy $g(\omega_j;d)$ of an oscillator with \(n\) energy levels can be deduced from Eq.\,(\ref{AydaHamiltonian}),
\begin{equation}
   E_{n,j} = \left(n + \frac{1}{2}\right) \hbar \omega_j(d), \quad n = 0, 1, 2, \ldots
   \label{GibbsOscillatorEnergy}
\end{equation}
obeying Bose-Einstein statistics and the average oscillator energy derived from quantum statistical mechanics can thus be written as,
\begin{equation}
   g(\omega_j) = \frac{\hbar \omega_j}{2} + \frac{\hbar \omega_j}{\exp\left(\frac{\hbar \omega_j}{kT}\right) - 1}.
   \label{GibbsOscillatorEnergybasedon1972BWN}
\end{equation}
This equation provides the basis for our subsequent analysis of the Casimir-Lifshitz interaction between planar surfaces. 
In their Nature paper from 1969, Parsegian and Ninham\,\cite{ParsegianNinham1969} used the harmonic oscillator model to re-derive the Lifshitz result for the case of three different planar dielectric media: $1|2|3$ with dielectric functions $\varepsilon_1(\omega)$, \,$\varepsilon_2(\omega)$,\,$\varepsilon_3(\omega)$, respectively. 
 Applying the relevant boundary conditions to the electromagnetic fields leads to the non-retarded dispersion equation for surface modes\,\cite{ParsegianNinham1969}, 
\begin{equation}
    D^{\text{NR}}(d, \omega) = 1 - \frac{[\varepsilon_1(\omega) - \varepsilon_2(\omega)] [\varepsilon_3(\omega) - \varepsilon_2(\omega)]}{[\varepsilon_1(\omega) + \varepsilon_2(\omega)] [\varepsilon_3(\omega) + \varepsilon_2(\omega)]} e^{-2 \kappa d}=0,
    \label{Eq_Ch1_sec:5:5}
\end{equation}
where, $\kappa$ is the wave vector. This dispersion equation is crucial for understanding the allowed electromagnetic surface modes in the system {which gives the complete solution to our problem.}  At zero temperature, the van der Waals-Casimir-Lifshitz interaction is simply the change in zero-point energies of the allowed quantized electromagnetic surface modes when two surfaces are separated by a finite distance, $d$ compared to when they are infinitely apart\,\cite{ParsegianNinham1969},
\begin{equation}
    E(d) = \frac{\hbar}{2} \sum_\lambda \int \frac{d^2 \kappa}{(2 \pi)^2} \left[ \omega_\lambda(d) - \omega_\lambda(\infty) \right],
    \label{Eq_Ch1_sec:5:5b}
\end{equation}
assuming an analytic function, $D$ with zeros at $\omega_\lambda (d)$, which has a derivative with singularities at $\omega_\lambda (\infty)$. Then the following equality is obtained after complex analysis\,\cite{NinhamParsegianWeiss1970},
\begin{equation}
    \sum_\lambda \frac{\hbar}{2} \left[ \omega_\lambda(d) - \omega_\lambda(\infty) \right] = \frac{1}{2 \pi i} \oint_C \frac{\hbar \omega}{2} \frac{d \omega}{D(d, \omega)} \frac{\partial D(d, \omega)}{\partial \omega},
    \label{Eq_Ch1_sec:5:6}
\end{equation}
where $C$ is the closed curve along the imaginary axis, where the closure is in the right-hand plane. All quantities vanish at the infinite semi-circle. Performing partial integration after substituting $i$$\xi$\ for $\omega$ \cite{NinhamParsegianWeiss1970},
\begin{equation}
\frac{1}{2 \pi i} \oint_C \frac{\hbar \omega}{2} \frac{d \omega}{D(d, \omega)} \frac{\partial D(d, \omega)}{\partial \omega} = \frac{\hbar}{4 \pi} \int_{-\infty}^\infty d\xi \, \ln \left[ D(d, i \xi) \right],
\label{Eq_Ch1_sec:5:7}
\end{equation}

\begin{equation}
E(d) \approx \frac{\hbar}{4 \pi^2} \int_0^\infty d\kappa \, \kappa \int_0^\infty d\xi \, \ln \left[ 1 - \Delta_{12}^{NR} \Delta_{32}^{NR} e^{-2 \kappa d} \right],
\label{Eq_Ch1_sec:5:8}
\end{equation}
where the non-retarded reflection coefficients are given as\,\cite{NinhamParsegianWeiss1970,Dzya},
\begin{equation}
\Delta_{ij}^{NR}=\frac{\varepsilon_i(i \xi)-\varepsilon_j(i \xi)}{\varepsilon_i(i \xi)+\varepsilon_j(i \xi)}.
\label{Eq_Ch1_sec:5:9}
\end{equation}
Accounting for the finite speed of light, equation (11) can be generalized to a retarded form by expressing it as a sum of transverse magnetic (TM) and transverse electric (TE) contributions\,\cite{NinhamParsegianWeiss1970}, 
\begin{equation}
E(d)\approx\frac{\hbar}{4 \pi^2} \int_0^\infty d\kappa \kappa\int_0^\infty d\xi \{ln [ 1-\vartheta(d)_{TM}]+ln [ 1-\vartheta(d)_{TE}]\}.
    \label{Eq_Ch1_sec:5:10}
\end{equation}
Having derived the expression for the retarded, zero-temperature case, we now consider the effect of finite temperature on the retarded Casimir-Lifshitz interaction. At non-zero temperatures, the zero-point energy of each mode is replaced by its corresponding Helmholtz free energy,\,\cite{NinhamParsegianWeiss1970} \newline $F(\omega,T)=k Tln[2 \sinh(\hbar \omega/[2 k T])]$.
 This leads to\,\cite{NinhamParsegianWeiss1970} 

\begin{equation}
\int_{-\infty}^\infty d \omega \frac{dF(\omega)}{d\omega} \ln[1-\vartheta].
    \label{Eq_Ch1_sec:5:12}
\end{equation}
The derivative of the Helmholtz free energy expression, obtained from partial integration,  provides a factor $coth[\hbar \omega/(2 k T))]-1$. The $coth$ factor has an infinite number of poles along the imaginary axis. This implies that zero and finite temperatures can be addressed by a simple substitution\,\cite{Dzya,NinhamParsegianWeiss1970},
\begin{equation}
\frac{\hbar}{2 \pi} \int_{0}^\infty d\xi \to kT \sum_{m=0}^\infty \,{}', 
\label{Eq_Ch1_sec:5:13}
\end{equation}
with $\xi_m = [2 \pi k T m]/\hbar$ and where the sum was originally from minus infinity to plus infinity leading to a factor of 1/2 for the \(m\)=0 term.

This leads us to the results obtained by Lifshitz and collaborators for a three-layer system where Casimir interactions occur across an intervening medium of vacuum or dilute gas with $\varepsilon_2(i \xi_m)=1$. The free energy of this finite-temperature, retarded interaction can be written as\,\cite{Dzya},
\begin{equation}
G(d,T) = \frac{k_B T}{2 \pi} \sum_{m=0}^\infty{}^\prime \int\limits_0^\infty d\kappa \, \kappa \sum_{\sigma} \ln \left( 1 - r_{\sigma}^{21} r_{\sigma}^{23} e^{-2\gamma_2 d} \right), 
\label{LifFreeEnergy}
\end{equation}
where $\sigma=\rm $TE,TM, and the prime in the sum indicates that the first term ($m$ = 0) is weighted by $1/2$. The Fresnel reflection coefficients between surfaces $i$ and $j$ for the TM, and TE polarizations are given by,
\begin{equation}
    r_{\rm TE}^{ij} = \frac{\gamma_i-\gamma_j}{\gamma_i+\gamma_j}\,;  \,\,\,\,\, r_{\rm TM}^{ij} = \frac{\varepsilon_j\gamma_i-\varepsilon_i \gamma_j}{\varepsilon_j \gamma_i+\varepsilon_i \gamma_j} \,. \label{eq:rtTETM}
\end{equation}
Here $\gamma_i= \sqrt{{\kappa}^2+\varepsilon_i\xi_m^2/c^2}$, with $i=1,2,3$ and the Matsubara frequency being $\xi_m=2 \pi k_B T m/\hbar$. In the limit of dilute media, the interaction leads to van der Waals interaction between the individual atoms\,\cite{PhysRevA.60.2581}.

\section{Results}
\label{Concepts}

\subsection{A Temperature-Distance Relation}
\label{TempDistOscillator}
At zero temperature, the {average energy for a given oscillator frequency}, described by Eq.\,(\ref{GibbsOscillatorEnergybasedon1972BWN}),  equals the zero point energy or ground state energy of the oscillator (this is to distinguish the zero point energy term being applied to the field). In the high-temperature limit, the oscillator energy is instead dominated by the thermal energy, $k T$.  At this stage of our analysis, we can provide a simple estimate for the link between temperature and distance based on the condition where the zero-point energy has the same magnitude as the thermal energy.
We note, based on the exponential factor in Eq.\,(\ref{LifFreeEnergy}), the frequencies that most contribute to the Lifshitz\,\cite{Dzya} free-energy are ${d\omega}/{c}\sim 1$ or less and determine the highest frequency contribution. 
This condition can be rewritten as ${\hbar\omega}/{2} \sim {\hbar c}/{2d}$, where the right-hand side represents the minimum energy fluctuation required for the interaction between two bodies separated by distance $d$. Meanwhile, the left-hand side corresponds to the mode's ground state energy. Suppose that the system is at a temperature such that $k T= {\hbar c}/{2d}$, then in the high temperature/small distance limit this happens to be the point where the thermal energy dominates the zero point energy for a single oscillator by Eq.\,(\ref{GibbsOscillatorEnergybasedon1972BWN}). We will later see the same relation come out from considerations of the zero-point energy term of the field and the radiation pressure. This can be important for a deeper appreciation of the \RevisedText{temperature-distance uncertainty relationship} ($\Delta(kT)\sim$$\hbar c / 2d$ discussed in Sec.\,\ref{HeisenbergOsc}).

\subsection{Casimir Critical Distance-Temperature Relation}


Similar temperature-distance relations are found for the Casimir effect in the low-temperature limit. To be specific, Ninham and Daicic derived a low-temperature expansion of the Casimir free energy between perfect metal plates that is also applicable for high-temperatures and small distances \,\cite{PhysRevA.57.1870},

\begin{equation}
G(d,T)\approx \frac{- \pi^2\ \hbar c}{720d^3}- \frac{\zeta(3) k^3 T^3}{2 \pi\hbar^2 c^2}+\frac{\pi^2 d k^4 T^4}{45\hbar^3 c^3} +..,
\label{BWNapproxFreeEnergy}
\end{equation}
where $\zeta(3)\approx1.202$ is a zeta function.
The first term is the attractive zero temperature Casimir result. The second term can be written as $\rho \hbar c/4\pi$, where $\rho$ is the density of photons\,\cite{LandauLifshitzStatPhys1} in blackbody radiation per unit volume and was interpreted by Ninham and Daicic as a chemical potential\,\cite{PhysRevA.57.1870}. The third is the black body radiation energy (at equilibrium) between the plates\,\cite{PhysRevA.57.1870}. Ninham {\it et al.} proposed that this repulsive black body radiation term exactly opposes the attractive Casimir term at equilibrium\,\cite{PhysRevA.67.030701,Ninham_Brevik_Bostrom_2022} and thus when $T={\hbar c}/{2d}$. For example, a temperature of 300\,K corresponds to $3.8\,\mu\,m$, and beyond this distance thermal effects dominate.
This is identical to the relation found by using the \RevisedText{temperature-distance uncertainty relation}. 
We also consider the transition to the high-temperature limit where the zero-frequency Matsubara term dominates, 
\begin{equation}
G_{m=0}=\frac{- 2 \zeta(3) k T}{ 16\pi d^2}.
\end{equation}
the cross over from the zero temperature contribution ($\frac{-\pi^{2}\hbar c}{720 d^{3}}$) occurs at,
\begin{equation}
 T=\frac{16 \pi^{3}}{720 \zeta(3)} \times \frac{\hbar c}{2 k d}\approx 0.57 \times \frac{\hbar c}{2 k d}\approx \frac{6.52 \times 10^{-4} \text{m K}}{d}.
\end{equation}
 Near room temperature, this corresponds to a distance of $2.3 \mu m$. For a pair of imperfect metal surfaces modeled by the Drude model, if the zero frequency transverse electric mode is assumed to be zero\,\cite{Bost2000}, the numerical prefactor changes ($0.57\rightarrow 1.14$).
 
Considering atom-atom interactions at large separations, we arrive at the zero temperature Casimir-Polder expression $V(d) = -[23 \hbar c \alpha^2(0)]/[4 \pi d^7]$.
Here $\alpha(0)$ is the static polarizability of the atoms, and the $d^{-7}$ indicates a notable weakening of the van der Waals interaction ($\propto d^{-6}$) with increasing separation. However, at finite temperatures and large separations ($d >> \hbar c/ k T$), the potential is dominated by a zero frequency term $V(d, T) = -[3 k_B T \alpha^2(0)]/[d^6]$\,\cite{PhysRevA.60.2581}.
The entropic term ($n = 0$ Matsubara frequency) dominates when these limiting results are equal and then leads to the distance-temperature relation, \begin{equation}
 T=\frac{23}{6 \pi} \times \frac{\hbar c}{2 k d}\approx 1.22 \times \frac{\hbar c}{2 k d}.
\end{equation}
This occurs at $d< \hbar c/ kT$ so we take the relation to be a rough approximation.

\subsection{A note on the thermal wavelength}
In a many body system, the  cross over from macroscopic classical to quantum behavior —exemplified by phenomena like superconductivity and Bose-Einstein condensation— occurs when the particle separations are less than or comparable to the thermal de-Broglie wavelength: $\lambda= \hbar/\sqrt{2\pi m k T}$. 
For purely academic interest, we substitute the Wick relation ($mc^{2}\sim \hbar c/2 d$) describing the range of virtual
excitations of a field of mass m, into 
thermal wavelength by using the mass of a real particle of the same field when $\lambda=d$, we arrive at an expression in the form of the temperature-distance relation:

\begin{equation}
{ T = \frac{h^2}{{2 m \pi k d^2}}\to \frac{2}{\pi} \frac{\hbar c}{{2 k d}}.}
    \label{kTHeisenbergUn2}
\end{equation}
While we equate the mass of a real particle to the corresponding property of a virtual excitation, the analysis remains within a single-particle framework and may offer further insights beyond the relationships arising from the Casimir effect.

\subsection{A Discussion of a Possible Semi-Classical Electromagnetic Contribution to Nuclear Physics}

We briefly discuss how the critical temperature for Casimir forces relates to the Ninham-Pask model for potential contributions to nuclear interactions.  As is well known from fundamental statistical physics\,\cite{landau2013statistical}, interactions between particles at high temperatures occur in the presence of a plasma of fluctuating, constantly created, and annihilated, electron-positron pairs. \RevisedText{Roberts and Butterfield\,\cite{Roberts_2020} urged caution in predicting that the time-energy uncertainty relation can induce virtual particles. Noting this caution, it is nonetheless compelling to consider the possibility that a bath of electron and positron particles may arise from quantum vacuum fluctuations between surfaces at nuclear distances. Following Landau and Lifshitz, it could potentially be useful to exploit the well-known relationship, valid at high temperatures, between temperature and e$^-$-e$^+$ plasma density\,\cite{LandauLifshitzStatPhys1}, 
\begin{equation}
\rho=\rho_{-} + \rho_{+}=\frac{3 \zeta(3) k^3 T^3}{ \pi^2\hbar^3 c^3}=\frac{3 \zeta(3)}{ 8 \pi^2 d^3}.
\label{eqn:7}
\end{equation}
 If such a plasma does arise, it may give rise to significant effects for interaction energies at the femtometer scale.} The interaction between nuclear particles follows a screened Yukawa potential and the question raised by Ninham {\it et al.} is whether the Casimir effect could contribute to these kinds of potentials. Surprisingly, similarities were found between screened Casimir forces in the presence of an electron-positron plasma and the nuclear interaction\,\cite{PhysRevA.67.030701,EPJDNinham2014,Ninham_Brevik_Bostrom_2022}.
\section{Discussion}
\label{Discussion}
\par 
Here, we have considered temperature-distance relations in various systems involving Casimir interactions. It is clear that there are striking resemblances of the several temperature-distance relations derived here with the energy-distance \RevisedText{uncertainty relation} derived by Wick from the energy-time \RevisedText{uncertainty relation}.
The relations found here have been deduced using arguments based on \RevisedText{the uncertainty relation} between thermal energy (temperature) and distance. Some insights can be sought from the fact that the exact relation emerges from equating the zero temperature Casimir term with a repulsive black body radiation term in an expansion of Casimir free energy between two perfect metal surfaces. 
In connection with this, we mention an earlier calculation\,\cite{PhysRevE.67.056116} of Casimir force between metal plates found a minimum when $kT \sim \hbar c/(2d)$, supporting our suggestion of a compensating effect between these terms. 
 \RevisedText{Finally, similar to Sec.\,\ref{TempDistOscillator} vacuum modes become discrete when bodies impose boundary conditions on the quantum fields. The mode frequencies depend on the separation distances of the bodies ($\omega \sim c/d$). When the system is at temperature $T$, it is in contact with a heat-bath so when
there is sufficient thermal energy available to excite a mode above the ground state, $k T \sim \hbar \omega \sim \hbar (c/d) \rightarrow T \sim \hbar c/kd$. } Notably, the critical distance-temperature corresponds to changes from quantum to classical behavior in interaction potentials at large separations and finite temperatures representing the correspondence principle in quantum systems\cite{PhysRevA.60.2581}.

\RevisedText{Here at the very end, we mention that the concept of a generalized uncertainty principle (GUP) has recently attracted considerable interest in other areas of physics, especially in gravitational theory in connection with the Casimir effect. The GUP serves to provide a new length scale, enabling one to revise the Heisenberg uncertainty principle to account for quantum gravitational effects at small scales.  Consider, for instance,  the discussion in E. Battista,  S. Capozziello and A. Errehymy\,\cite{Battista2024} and references therein.}


\section*{Acknowledgments}

This research is part of the project No. 2022/47/P/ST3/01236 co-funded by the National Science Centre and the European Union's Horizon 2020 research and innovation programme under the Marie Sk{\l}odowska-Curie grant agreement No. 945339. The research by MB and JJM took place at the "ENSEMBLE3-Center of Excellence for nanophononics, advanced materials and novel crystal growth-based technologies" project (GA No. MAB/2020/14) carried out under the International Research Agenda programs of the Foundation for Polish Science that are co-financed by the European Union under the European Regional Development Fund and the European Union Horizon 2020 research and innovation program Teaming for Excellence (GA. No. 857543) for supporting this work. 
MB and JJM's research contributions to this publication were created as part of the project of the Minister of Science and Higher Education "Support for the activities of Centers of Excellence established in Poland under the Horizon 2020 program" under contract No. MEiN/2023/DIR/3797. Two unknown referees are gratefully acknowledged for the useful references and interpretations.


\end{document}